\documentclass{pspum-l}

\usepackage{}

\usepackage{dsfont} 
\usepackage{physics}
\usepackage{tikz} 
\usetikzlibrary{arrows} 

\usepackage{cleveref}

\newcommand{\SVseven}{\text{SV}^{\text{G}_2}}
\newcommand{\SVeight}{\text{SV}^\text{Spin(7)}}

\newcommand{\normord}[1]{:\mathrel{#1}:} 

\newcommand*\Sc{\mathbb{S}}
\newcommand*\R{\mathbb{R}}

\copyrightinfo{}{}

\theoremstyle{definition}

\theoremstyle{remark}

\numberwithin{equation}{section}

\begin{document}

\title[Superconformal algebras for the Schoen Calabi--Yau manifold]{Superconformal algebras for the\\
Schoen Calabi--Yau manifold}


\author{Mateo Galdeano}
\address{Fachbereich Mathematik, Universit\"at Hamburg, Bundesstra{\ss}e 55, 20146, Hamburg, Germany}
\email{mateo.galdeano@uni-hamburg.de}
\thanks{This work was supported by a scholarship from the Mathematical Institute, University of Oxford, as well as the Deutsche Forschungsgemeinschaft (DFG, German Research Foundation) under Germany’s Excellence Strategy – EXC 2121 ``Quantum Universe'' – 390833306}

\subjclass[2020]{Primary 81R10, 81Q70;
                Secondary 81T30, 81T33}

\date{26/06/2023}

\begin{abstract}
We revisit the proposal of \cite{Fiset:2021ruv} for the worldsheet description of string theory compactifications on special holonomy manifolds obtained via connected sums: the geometric construction corresponds to a diamond of inclusions of worldsheet algebras. We present new  evidence for the proposal by considering compactifications on the Schoen Calabi--Yau manifold.
\end{abstract}

\maketitle


\tableofcontents

\section{Introduction}

The purpose of this note is to provide further evidence in favour of a proposal of \cite{Fiset:2021ruv} regarding the worldsheet formulation of string theory compactifications on connected sum manifolds.

Connected sum manifolds are obtained by gluing two Asymptotically Cylindrical (ACyl) open manifolds along isomorphic asymptotic ends. This technique has proved extremely fruitful to construct compact manifolds of special holonomy \cite{Kovalev:2003, Corti:2012, Corti:2012kd, Crowley:2015ctv, Nordstrom:2018cli, Goette:2020, Braun:2018joh} which have then been used to study string compactifications, see for example \cite{Halverson:2014tya, Braun:2017ryx, Braun:2017uku, Braun:2018vhk, Heckman:2018mxl} and references therein.

From the point of view of the worldsheet, a string theory compactification on a special holonomy manifold corresponds to an extended algebra of chiral symmetries \cite{Odake:1988bh, Shatashvili:1994zw, Figueroa-OFarrill:1996tnk}. This can be understood from the fact that covariantly constant forms give rise to new worldsheet symmetries and their associated currents \cite{Howe:1991vs, Howe:1991im,  Howe:1991ic}.

This correspondence is further enhanced when the compactifying manifold can be described in terms of a connected sum. The features of this geometric structure are reflected in the worldsheet in terms of a diamond of inclusions of chiral algebras \cite{Fiset:2021ruv}. This has been verified for Twisted Connected Sum (TCS) \cite{Fiset:2018huv}, Extra Twisted Connected Sum (ETCS) and Generalized Connected Sum (GCS) manifolds \cite{Fiset:2021ruv}.

In this note we check that this proposal also holds for the Schoen Calabi--Yau manifold \cite{Schoen1988}. The connected sum structure of this manifold was obtained in \cite{Braun:2017uku} by applying M-theory/heterotic duality to TCS G$_2$-manifold compactifications. Consequently, we will find that the associated diamond of algebras presents some similarities with that of the TCS.

This note is organised as follows: we first introduce in \cref{sec:Background} some necessary background on connected sums and worldsheet algebras together with a summary of the proposal of \cite{Fiset:2021ruv}. In \cref{sec:TCS} we discuss as a warm-up case---and with the purpose of setting up notation---the example of TCS G$_2$-manifolds. This case was already analysed in \cite{Fiset:2018huv}, but here we have modified the presentation to follow \cite{Fiset:2021ruv}.

\Cref{sec:Schoen} contains the most important results of this note. We begin with a review of the connected sum structure of the Schoen manifold following \cite{Braun:2017uku} before introducing the corresponding diamond of chiral algebras. We proceed to study the diamond in detail, looking at the the relationship between the algebra at the bottom and the intersection of the algebras at the lateral tips. We then study automorphisms of the diamond and relate them to mirror symmetry maps, before summarizing our results and concluding in \cref{sec:Conclusion}.

\section{Background and general discussion}

\label{sec:Background}

\subsection{Special holonomy manifolds via connected sums}

\label{sec:introConnSum}

The Twisted Connected Sum (TCS) construction was introduced by Kovalev \cite{Kovalev:2003} and later extended by Corti, Haskins, Nordstr\"om and Pacini \cite{Corti:2012, Corti:2012kd} to provide new examples of compact manifolds with G$_2$-holonomy. This technique has been further generalized within the G$_2$-holonomy setting to the Extra Twisted Connected Sum (ETCS) construction \cite{Crowley:2015ctv, Nordstrom:2018cli, Goette:2020}, and several alternative versions of these ``connected sum'' constructions have been proposed with the aim of producing new manifolds with holonomy $\text{SU}(3)$ \cite{Doi:2013}, $\text{SU}(4)$ \cite{Doi:2015} or $\text{Spin}(7)$ \cite{Braun:2018joh, Doi:2019}.\footnote{The Generalized Connected Sum (GCS) of Spin(7)-holonomy manifolds proposed in \cite{Braun:2018joh} is supported by strong evidence from string theory compactifications \cite{Braun:2018joh, Cvetic:2020piw, Fiset:2021ruv}. Nevertheless, a formal mathematical proof of its general validity is still missing in the literature.}

We now describe the general ideas underlying all these constructions, and refer the reader to the original papers as well as \cite{Langlais:2023pmo} for the technical details. First of all, we introduce some basic definitions. We say a manifold $M$ \emph{asymptotes} the cylinder $\R^+\times N$ if there exists a compact submanifold of $M$ whose complement is diffeomorphic to such cylinder. In that case, we call $M$ an \emph{Asymptotically Cylindrical} (ACyl) manifold with \emph{cross-section} $N$ and we represent the asymptotic behaviour with an arrow
\begin{equation}
    M\longrightarrow\R^+\times N \, .
\end{equation}
Furthermore, we require the metric and certain covariantly constant forms on $M$---which will be determined by the holonomy of the ACyl manifold---to asymptote those of $\R^+\times N$ at a certain rate.  The technical details of the analysis involved in this asymptotic behaviour will not be relevant for us, hence we omit them. We parametrize the $\R^+$ factor by a coordinate $t$ and represent the asymptotic behaviour of the tensors by an arrow. In particular, for the metric we require
\begin{equation}
    g_M \longrightarrow g_{M,\infty}=\dd t^2+g_N \, ,
\end{equation}
where the subscript $\infty$ refers to the fact that the equality holds in the limit $t\longrightarrow\infty$.

We want to obtain a compact manifold $M$ with holonomy equal to $G$. To this end, we consider two ACyl manifolds $M_\pm$ with cross-sections $N_\pm\,$. Both $M_\pm$ must have holonomy contained in $G$, so they are both equipped with torsion-free $G$-structures. We also require the cross-sections $N_\pm$ to be isomorphic. We call these ACyl manifolds the \emph{building blocks} of the construction.

The next step is to glue the two building blocks along their isomorphic asymptotic ends to produce a compact manifold $M$. Parametrizing the directions of the cylinders $\R^+_\pm$ by $t_\pm\,$, we truncate both cylinders and introduce boundaries at $t_\pm=t_0+1$. We can then define an isomorphism $F_{t_0}$ between $\left[ t_0,t_0+1 \right]_+\times N_+$ and $\left[ t_0,t_0+1 \right]_-\times N_-$ as follows
\begin{equation}
    F_{t_0}:\left[ t_0,t_0+1 \right]_+\times N_+ \ni \left(t,n\right) \longmapsto \left(2 t_0+1-t,\phi(n)\right) \in \left[ t_0,t_0+1 \right]_-\times N_- \, ,
\end{equation}
where $\phi$ is an isomorphism between $N_+$ and $N_-\,$. We call $F_{t_0}$ the \emph{gluing map}, it identifies the asymptotic ends of $M_\pm$ and produces a compact manifold $M$. The area where the gluing is performed is usually denominated the \emph{neck} region of $M$, and we will denote it by $M_+\cap M_-\,$. The gluing map is chosen such that the $G$-structures of $M_\pm$ can be patched together into a globally well-defined $G$-structure on $M$.

As we have truncated $M_\pm$ at $t_\pm=t_0+1$, the $G$-structure on $M$ has non-trivial torsion localized in the neck region. The key point of the construction is that through some careful analysis it can be shown that for a sufficiently long neck---equivalently, for large enough $t_0$---it is possible to deform the $G$-structure to a torsion-free one. This completes the construction. Once again, the technical aspects of the analysis are unimportant for our purposes and we will not discuss them.

\subsection{Worldsheet algebras for special holonomy manifolds}

\label{sec:introAlgebras}

We now focus on worldsheet aspects of string theory compactifications on special holonomy manifolds. For the purpose of this note, we will consider supersymmetric type II backgrounds given by a direct product of a special holonomy manifold $M$ and Minkowski spacetime. In practice, we will ignore the spacetime factor and study these compactifications in terms of an $\mathcal{N}=(1,1)$ non-linear sigma model with target $M$. From the worldsheet point of view, this is described by a two-dimensional chiral superconformal field theory. Throughout this note we restrict ourselves to one of the chiral components, say the holomorphic one, and study the underlying superconformal algebra.

The special holonomy of the target manifold $M$ has important consequences for the sigma model. Recall that reduced holonomy can be equivalently formulated as the existence of certain covariantly constant forms on the manifold \cite{Joyce:2007}. It was observed by Howe and Papadopoulos \cite{Howe:1991vs, Howe:1991im,  Howe:1991ic} that each of these covariantly constant forms leads to an additional (classical) non-linear chiral symmetry on the sigma model action. The conserved (super-)currents associated to these symmetries are constructed directly from the covariantly constant forms.

At the level of the quantum worldsheet theory, this implies the existence of additional chiral operators corresponding to the covariantly constant forms on $M$. These operators come in supersymmetric pairs: to each $p$-form we can associate an operator of conformal weight $\frac{p}{2}$ and its supersymmetric partner of weight $\frac{p+1}{2}\,$. Furthermore, these operators retain some features reminiscent of their geometric origin, as we will motivate in more detail in an example below. 

We thus find that the original underlying $\mathcal{N}=1$ (super-)Virasoro algebra, given by the (holomorphic) stress-tensor $T$ and the chiral supersymmetry current $G$, can be extended by additional operators to obtain a new $\mathcal{N}=1$ W-algebra \cite{Odake:1988bh, Shatashvili:1994zw, Figueroa-OFarrill:1990tqt, Figueroa-OFarrill:1990mzn, Blumenhagen:1991nm, Figueroa-OFarrill:1996tnk}. This means that we can associate to each special holonomy manifold $M$ a particular chiral algebra $\mathcal{A}(M)$ that captures the underlying chiral symmetries upon compactification.

We illustrate this with an example: consider a K\"ahler $n$-fold, that is, a $2n$-dimensional manifold with $\text{U}(n)$-holonomy. This manifold comes equipped with a covariantly constant Hermitian 2-form $\omega_n\,$, which gives rise to a weight $1$ superprimary multiplet $(J^3_n,G^3_n)$. This extends the original $\mathcal{N}=1$ Virasoro algebra to an $\mathcal{N}=2$ Virasoro algebra and reflects the fact that a compactification on a K\"ahler manifold preserves $\mathcal{N}=2$ supersymmetry.

The most relevant case for us will be when the holonomy is further reduced to $\text{SU}(n)$, which corresponds to a Calabi--Yau $n$-fold. In this situation we have an additional covariantly constant complex $n$-form known as the holomorphic volume form $\Omega_n\,$. This results in two additional real superprimary multiplets of conformal weight $\frac{n}{2}$ corresponding to the real and imaginary part of $\Omega_n\,$. We denote them by $(A_n,C_n)$
and $(B_n,D_n)$, respectively. These further extend the $\mathcal{N}=2$ Virasoro algebra of the K\"ahler case to the \emph{Odake algebra} \cite{Odake:1988bh}, which we denote by $\text{Od}_n \,$.
 
For $n\geq 3$ the Odake algebra is associative modulo the following \emph{singular} fields and their descendants (which we call \emph{null} fields)
\begin{equation}
\label{eq:nullfields}
N^1_n = \partial A_n \; -\normord{J^3_n\, B_n} \, , \qquad
N^2_n = \partial B_n \; +\normord{J^3_n\, A_n} \, ,
\end{equation}
where colons represent normal ordering. Singular fields are descendants which are annihilated by all positive modes of the generators of the algebra. They have zero norm, so they do not contribute to correlation functions and they must be quotiented out of our extended algebras. This is a subtle point in the analysis of this note, as many of the statements we check are true ``up to null fields''.

The case $n=2$ of the Odake algebra is very special as it corresponds to K3 surfaces, which are also hyper-K\"ahler manifolds. Some of the geometric features of these manifolds are indeed present in $\text{Od}_2\,$: recall there exists a triple of complex structures $I$, $J$, $K$ on the K3 surface satisfying the identities
\begin{equation}
\label{eq:hypercomplexrelations}
I^2=J^2=K^2=-\text{Id}\,, \qquad I J=-J I=K.
\end{equation}
Each of them has a corresponding Hermitian form $\lbrace \omega^I,\omega^J,\omega^K \rbrace$, and these give rise to the operators $\lbrace J^3_2, A_2, B_2 \rbrace$.\footnote{Note the $\text{SU}(2)$ forms are given in terms of the Hermitian forms by
\begin{equation*}
    \omega_2=\omega^I\, , \qquad \Omega_2=\omega^J+i\,\omega^K \, .
\end{equation*}
} In addition to \eqref{eq:nullfields} the following fields are also singular
\begin{equation}
    N^3_2=\,\normord{J^3_2 J^3_2} - \normord{A_2 A_2} \, , \qquad N^4_2=\,\normord{A_2 A_2} - \normord{B_2 B_2} \, ,
\end{equation}
this is a manifestation of the first identity of \eqref{eq:hypercomplexrelations} at the algebra level. Another example of the correlation between geometry and algebra can be seen as follows: the $\text{Od}_2$ algebra is actually isomorphic to the (little) $\mathcal{N}=4$ Virasoro algebra, obtained by extending $\mathcal{N}=1$ Virasoro by an $\text{SU}(2)$ Kac--Moody current super-algebra at level $k=1$. This is in direct correspondence with the $\text{SU}(2)$-structure of the K3 surface, and indeed the OPEs of the Kac--Moody current reproduce---in an appropriate sense---the second identity of \eqref{eq:hypercomplexrelations}.

We conclude this section mentioning some additional W-algebras that will be relevant for the discussion. The extended algebras associated to manifolds of holonomy G$_2$ and Spin(7) were introduced by Shatashvili and Vafa \cite{Shatashvili:1994zw}. We call these the Shatashvili--Vafa G$_2$ and Spin(7) algebras, and denote them by $\SVseven$ and $\SVeight$ respectively. The $\SVseven$ algebra has a superprimary multiplet $(P,K)$ of weight $\frac{3}{2}$ corresponding to the \emph{associative} 3-form $\varphi$ and a multiplet $(X,M)$ of weight $2$ corresponding to the \emph{coassociative} 4-form $\psi$. We will not make use of the $\SVeight$ algebra, but we mention that it has a multiplet of weight $2$ corresponding to the \emph{Cayley} 4-form $\Psi$.

Finally, when the target manifold is either $\R$ or $\Sc^1$---which corresponds to the case of trivial holonomy---we can parametrize it by a coordinate $t$. The form $\dd t$ is covariantly constant and the associated currents are a free Majorana--Weyl fermion $\psi_t$ and a $\widehat{\mathfrak{u}}(1)$ current $j_t=i\partial_t \,$. These generate what we call the \emph{Free algebra}, denoted by $\text{Fr}^1$. The $\mathcal{N}=1$ Virasoro algebra can be found as a subalgebra of $\text{Fr}^1$, with its generators given by
\begin{equation} 
G^t = \, \normord{j_t\psi_t} \, , \qquad
T^t = \frac{1}{2}\normord{\left(\partial\psi_t\psi_t + j_t j_t\right)} .
\end{equation}
We will keep reusing this notation to describe the Virasoro algebra inside $\text{Fr}^1$. When $n$ copies of $\R$ or $\Sc^1$ are considered, we obtain $n$ copies of the Free algebra $\text{Fr}^n=\text{Fr}^1\oplus\overset{n}{\cdots}\oplus\text{Fr}^1$.

\subsection{Worldsheet algebras for connected sum manifolds}

\label{sec:introConnSumAlgebras}

When the manifold $M$ has a connected sum structure, the correspondence between the holonomy of $M$ and its associated W-algebra---which we denote by $\mathcal{A}(M)$---can be further extended. It has been proposed \cite{Fiset:2018huv, Fiset:2021ruv} that the piecewise decomposition of $M$ has its counterpart at the level of worldsheet superconformal algebras, so that chiral algebras can be associated to the building blocks $M_\pm$ as well as the neck region $M_+\cap M_-$ and arranged into a diamond of inclusions that matches the embeddings in geometry.

The argument for the proposal goes as follows: we consider an open submanifold $U$ of the manifold $M$ and study the algebra $\mathcal{A}(U)$ of chiral operators associated to $U$. The covariantly constant forms on $M$ restrict to covariantly constant forms on $U$, so the corresponding operators also belong to $\mathcal{A}(U)$. Nevertheless, there can be covariantly constant forms on $U$ which are not well-defined over the whole of $M$. Equivalently, the holonomy of $U$ might be further reduced than the holonomy of $M$. This means the chiral algebras and the corresponding submanifolds follow the opposite inclusion pattern
\begin{equation}
\label{eq:subalgebraembedding}
U\subset M \qquad \Leftrightarrow \qquad \mathcal{A}(U)\supset \mathcal{A}(M) \, .
\end{equation}
We point out that the algebra inclusion in \eqref{eq:subalgebraembedding} might not be unique, reflecting the additional geometric freedom available in the submanifold $U$. Consider now a different open submanifold $V\subset M$, which again satisfies $\mathcal{A}(V)\supset \mathcal{A}(M)$, and the overlap of the two open patches $U\cap V$. Following the reasoning of \eqref{eq:subalgebraembedding}, we can find $\mathcal{A}(M)$ as a subalgebra of $\mathcal{A}(U\cap V)$ via the inclusion $\mathcal{A}(U)\subset \mathcal{A}(U\cap V)$ or the inclusion $\mathcal{A}(V)\subset \mathcal{A}(U\cap V)$. Since $\mathcal{A}(M)$ represents the chiral symmetries of the whole manifold, both realizations of the inclusion should agree: this restricts some of the freedom present in the embedding \eqref{eq:subalgebraembedding}. The same argument can be repeated with further open patches of $M$ and in the end the embedding of $\mathcal{A}(M)$ should be uniquely fixed, as there should be no ambiguity in identifying the chiral symmetries of the theory.

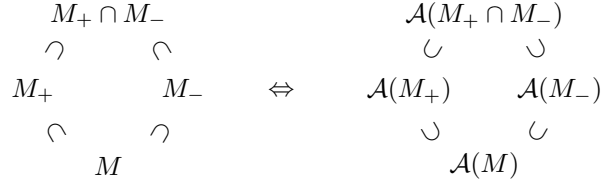
\begin{figure}
\begin{tikzpicture}
\node at (4.3,0.55) {\rotatebox{240}{$\subset$}};
\node at (5.7,0.55) {\rotatebox{300}{$\subset$}};
\node at (4.3,-0.55) {\rotatebox{300}{$\subset$}};
\node at (5.7,-0.55) {\rotatebox{240}{$\subset$}};
\node at (5,1) {$M_+\cap M_-$};
\node at (4,0) {$M_+$};
\node at (6,0) {$M_-$};
\node at (5,-1) {$M$};
\node at (7.3,0) {$\Leftrightarrow$};
\node at (9.3,0.55) {\rotatebox{60}{$\subset$}};
\node at (10.7,0.55) {\rotatebox{120}{$\subset$}};
\node at (9.3,-0.55) {\rotatebox{120}{$\subset$}};
\node at (10.7,-0.55) {\rotatebox{60}{$\subset$}};
\node at (10,1) {$\mathcal{A}(M_+\cap M_-)$};
\node at (9,0) {$\mathcal{A}(M_+)$};
\node at (11,0) {$\mathcal{A}(M_-)$};
\node at (10,-1) {$\mathcal{A}(M)$};
\end{tikzpicture}
\caption{Diamond of submanifold embeddings and diamond of subalgebra inclusions for a generic connected sum manifold.}
\label{fig:Diamond}
\end{figure}

Connected sum manifolds constitute a very particular case since they are completely described by just two open submanifolds---the building blocks $M_\pm$---glued along their intersection---the neck region $M_+\cap M_-\,$. Following the previous arguments, we should be able to arrange the chiral algebras associated to $M$, $M_\pm$ and $M_+\cap M_-$ in a diamond of inclusions uniquely determined by the geometry of the connected sum construction. We display this in \cref{fig:Diamond}.

This line of reasoning is a motivation rather than a proof, and the validity of this proposal has to be checked on a case-by-case basis. This has been done so far for TCS G$_2$-manifolds \cite{Fiset:2018huv}, ETCS G$_2$-manifolds and GCS Spin(7)-manifolds \cite{Fiset:2021ruv}. We present in \cref{sec:Schoendiamond} a new check of the proposal for the Schoen manifold.

The gluing map of a connected sum is an isomorphism of $M_+\cap M_-$ that provides a well-defined $G$-structure on the compact manifold $M$. An analogous structure should also be present for the diamond and this is indeed the case: the gluing map can be translated into an automorphism of $\mathcal{A}(M_+\cap M_-)$ that preserves $\mathcal{A}(M)$.

We now comment on the chiral algebra corresponding to the intersection of the lateral tips of the diamond $\mathcal{A}(M_+)\cap\mathcal{A}(M_-)$. By construction, $\mathcal{A}(M)$ is contained inside this algebra and we might wonder if this inclusion is strict. If that was the case, this would mean that the set of chiral symmetries of $M$ could be further enhanced beyond those associated to a generic manifold with $G$-holonomy. On the other hand, an agreement between these algebras would indicate that the connected sum construction considered produces---at least from a worldsheet point of view---generic manifolds of $G$-holonomy. An analytical proof of this equality is currently beyond reach as the vacuum characters of extended W-algebra intersections are poorly understood. Nevertheless, we provided in \cite{Fiset:2021ruv} numerical checks for the connected sums mentioned above up to a certain level in the vacuum modules. Similar checks are presented in this note for the Schoen manifold.

Finally, we would like to point out that the diamond of algebras can be used to study mirror symmetry for connected sum manifolds $M$. A mirror map in geometry corresponds to a certain algebra automorphism of the associated chiral algebra $\mathcal{A}(M)$. However, if $M$ admits a connected sum decomposition, a mirror map respecting the connected sum structure should correspond to an automorphism of the whole diamond of inclusions. This was studied in \cite{Fiset:2021ruv} for GCS Spin(7)-manifolds. Even though the only automorphism of $\SVeight$ is the identity, it was found that non-trivial automorphisms of the whole diamond exist and these were used to propose new GCS mirror maps beyond the ones described in \cite{Braun:2019lnn}. The validity of these mirror maps was explicitly checked for GCS Spin(7)-manifolds given by Joyce orbifolds \cite{Joyce:1996} admitting an SYZ fibration \cite{Strominger:1996it}.

\section{Review of superconformal algebras for TCS G$_2$-manifolds}

\label{sec:TCS}

In this section we briefly recall the Twisted Connected Sum construction (TCS) of G$_2$-holonomy manifolds \cite{Kovalev:2003, Corti:2012, Corti:2012kd} and the corresponding diamond of superconformal algebras \cite{Fiset:2018huv, Fiset:2021ruv}. This serves the twofold purpose of establishing some notation that will be used throughout the rest of the note as well as presenting the construction of the TCS diamond as a warm-up case for the Schoen manifold.

\subsection{Geometry of the TCS construction}

\label{sec:TCSgeometry}

As discussed in \cref{sec:introConnSum}, the idea of the TCS construction is to glue two Asymptotically Cylindrical (ACyl) manifolds---called \emph{building blocks}---along their isomorphic asymptotic ends. This is done via a \emph{gluing map} that identifies the \emph{cross-sections} of the cylinders and produces a compact manifold $M$ with a long \emph{neck} region. This manifold has a well-defined G$_2$-structure that becomes torsion-free after a certain deformation.

Firstly, we need to introduce the concept of Asymptotically Cylindrical (ACyl) Calabi--Yau $n$-fold (CY$_n$). This is an open Calabi--Yau $n$-fold which asymptotes a cylinder with cross-section given by a Calabi--Yau $(n-1)$-fold times a circle $\Sc^1$, that is
\begin{equation}
    \text{ACyl CY}_n\longrightarrow\R^+\times \Sc^1\times\text{CY}_{n-1} \, .
\end{equation}
Denoting by $(t,\theta)$ the coordinates of $\R^+\times \Sc^1$, the metric $g_n\,$, Hermitian form $\omega_n$ and holomorphic volume form $\Omega_n$ of the $\text{ACyl CY}_n$ satisfy the following asymptotic relations
\begin{align}
\label{eq:metricCYasymptotic}
    g_n & \longrightarrow g_{n,\infty}=\dd t^2+\dd\theta^2+g_{n-1} \, ,\\
    \label{eq:hermitianasymptotic}
    \omega_n & \longrightarrow \omega_{n,\infty}=\dd t\wedge\dd\theta + \omega_{n-1} \, ,\\
    \label{eq:holomorphicasymptotic}
    \Omega_n & \longrightarrow \Omega_{n,\infty}=(\dd\theta-i\dd t)\wedge\Omega_{n-1} \, ,
\end{align}
with the notation introduced in \cref{sec:introConnSum}.

For a TCS manifold, each of the building blocks consist on an ACyl Calabi--Yau 3-fold times a circle, $M_\pm=\Sc^1_{\xi\pm}\times\text{ACyl CY}_{3\pm}\,$. We call these circles the \emph{external} circles and parametrize them by coordinates $\xi_\pm\,$. The cross-sections of these building blocks are given by $N_\pm=\text{CY}_{2\pm}\times\Sc^1_{\theta\pm}\times\Sc^1_{\xi\pm}$, where the circles $\Sc^1_{\theta\pm}$ coming from the $\text{CY}_{3\pm}$ are called \emph{internal} circles and we denote their coordinates by $\theta_\pm\,$. Thus, the building blocks satisfy
\begin{equation}
\label{eq:TCSasymptoticbuildingblock}
    M_\pm=\Sc^1_{\xi\pm}\times\text{ACyl CY}_{3\pm}\longrightarrow\R^+_\pm\times\Sc^1_{\theta\pm}\times\Sc^1_{\xi\pm}\times\text{CY}_{2\pm}= \R^+_\pm\times N_\pm \, .
\end{equation}
Each of the building blocks admits a G$_2$-structure that can be described in terms of an \emph{associative} 3-form. We denote these forms by $\varphi_\pm$ and define them in terms of the forms on the Calabi--Yau 3-folds and the external circles
\begin{equation}
\label{eq:defassociative}
    \varphi_\pm=\dd\xi_\pm\wedge\omega_{3\pm}+\Re(\Omega_{3\pm})\, .
\end{equation}
Parametrizing the $\R^+_\pm$ factors by coordinates $t_\pm \,$, these forms follow the asymptotic behaviour
\begin{equation}
\label{eq:coassociativeasymptotic}
    \varphi_\pm\longrightarrow \dd\xi_\pm\wedge\omega^I_{\pm}+\dd\theta_\pm\wedge\omega^J_{\pm}+\dd t_\pm\wedge\omega^K_{\pm}+\dd\xi_\pm\wedge\dd t_\pm\wedge\dd\theta_\pm\, ,
\end{equation}
where we have made explicit the hyper-K\"ahler structure of the K3 surfaces $\text{CY}_{2\pm}$ identifying
\begin{equation}
    \omega_{2\pm}=\omega^I_\pm\, , \qquad \Omega_{2\pm}=\omega^J_\pm+i\,\omega^K_\pm \, .
\end{equation}
The gluing map $F$ between the two asymptotic ends identifies the external circle on $M_+$ with the internal circle on $M_-$ and vice-versa. In order to compensate this ``twist'', the K3 surfaces are identified via a \emph{hyper-K\"ahler matching}: this is an isometry of the K3 surfaces that effectively rotates the hyper-K\"ahler structures. At the level of differential forms, the effect of the gluing map $F$ is as follows
\begin{align}
\label{eq:TCSgluingmap1}
    F^*(\dd t_-)&=-\dd t_+\, ,& F^*(\dd\xi_-)&=\dd\theta_+\, ,& F^*(\dd\theta_-)&=\dd\xi_+\, , \\
    \label{eq:TCSgluingmap2}
    F^*(\omega^K_-)&=-\omega^K_+\, ,& F^*(\omega^I_-)&=\omega^J_+\, ,& F^*(\omega^J_-)&=\omega^I_+\, .
\end{align}
Note the gluing map identifies the asymptotic ends of the associative 3-forms $\varphi_\pm$ \eqref{eq:coassociativeasymptotic} and provides a well-defined G$_2$-structure to the compact manifold $M$. We denote the corresponding associative 3-form by $\varphi$.

\subsection{Chiral algebra perspective}

\label{sec:TCSalgebra}

We now summarize how the diamond of algebras corresponding to the TCS construction can be obtained, showing explicitly the inclusions of the most important generators. The guiding principle for the algebra inclusions are the asymptotic geometric relations we have just described. The algebra at the top of the diamond corresponds to the neck region, which has the form $\R^+\times\Sc^1\times\Sc^1\times\text{CY}_{2}\,$. This immediately implies that the associated algebra is $\text{Fr}^3\oplus \text{Od}_2\,$. The neck region can be described from the point of view of $R^+_+\times N_+$ or $R^+_-\times N_-$, to be completely explicit we follow the $R^+_+\times N_+$ description here.

\begin{figure}
\begin{tikzpicture}
\node at (2.8,0.55) {\rotatebox{240}{$\subset$}};
\node at (4.2,0.55) {\rotatebox{300}{$\subset$}};
\node at (2.8,-0.55) {\rotatebox{300}{$\subset$}};
\node at (4.2,-0.55) {\rotatebox{240}{$\subset$}};
\node at (3.5,1) {$\R^+\times\Sc^1_\theta\times\Sc^1_\xi\times\text{CY}_{2}$};
\node at (2,0) {$\Sc^1_{\xi+}\times\text{CY}_{3+}$};
\node at (5,0) {$\Sc^1_{\xi-}\times\text{CY}_{3-}$};
\node at (3.5,-1) {$\text{TCS G}_2$};
\node at (6.8,0) {$\Leftrightarrow$};
\node at (9.3,0.55) {\rotatebox{60}{$\subset$}};
\node at (10.7,0.55) {\rotatebox{120}{$\subset$}};
\node at (9.3,-0.55) {\rotatebox{120}{$\subset$}};
\node at (10.7,-0.55) {\rotatebox{60}{$\subset$}};
\node at (10,1) {$\text{Fr}^3\oplus \text{Od}_2$};
\node at (8.8,0) {$\left( \text{Fr}^1\oplus \text{Od}_3 \right)_+$};
\node at (11.3,0) {$\left( \text{Fr}^1\oplus \text{Od}_3 \right)_-$};
\node at (10,-1) {$\SVseven$};
\end{tikzpicture}
\caption{Diamond of submanifold embeddings and diamond of subalgebra inclusions for a TCS G$_2$ manifold.}
\label{fig:DiamondTCS}
\end{figure}
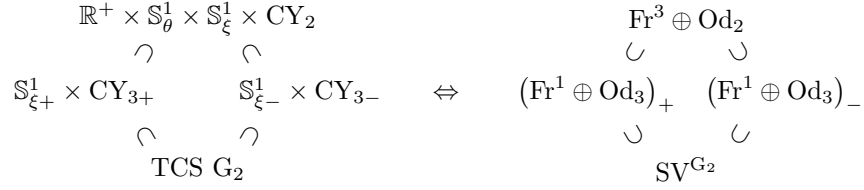

The lateral tips of the diamond correspond to the building blocks $M_\pm=\Sc^1_{\xi\pm}\times\text{ACyl CY}_{3\pm}$ so we have to associate algebras $\left( \text{Fr}^1\oplus \text{Od}_3 \right)_\pm$ to them. The right ansatz for the inclusion $\left( \text{Fr}^1\oplus \text{Od}_3 \right)_+\subset\text{Fr}^3\oplus \text{Od}_2$ is provided by the asymptotic behaviour of the forms \eqref{eq:hermitianasymptotic}, \eqref{eq:holomorphicasymptotic}, which translates into the language of chiral operators as
\begin{equation}
\label{eq:Od3fromOd2}
    J^3_{3+} = J^3_2 \, + \normord{\psi_t\,\psi_\theta} \, , \qquad A_{3+}+iB_{3+} = \, \normord{(\psi_\theta-i\psi_t)(A_2+iB_2)} \, .
\end{equation}
Since $M$ has G$_2$-holonomy, the algebra at the bottom of the diamond must be $\SVseven$. The inclusion $\SVseven\subset\left( \text{Fr}^1\oplus \text{Od}_3 \right)_+$ was already studied in \cite{Figueroa-OFarrill:1996tnk}, and the relation \eqref{eq:defassociative} provides the correct ansatz for our presentation
\begin{equation}
    P = \, \normord{\psi_\xi\,J^3_3} + A_3 \, .
\end{equation}
Combining this with \eqref{eq:Od3fromOd2} we obtain the ansatz for the inclusion $\SVseven\subset\text{Fr}^3\oplus \text{Od}_2\,$, which matches the expectation from \eqref{eq:coassociativeasymptotic}
\begin{equation}
    P = \, \normord{\psi_\xi J^3_2} + \normord{\psi_\theta\, A_2} +\normord{\psi_t\, B_2} + \normord{\psi_\xi\,\psi_t\,\psi_\theta} .
\end{equation}
The TCS gluing map \eqref{eq:TCSgluingmap1}, \eqref{eq:TCSgluingmap2} naturally provides an ansatz for an automorphism of $\text{Fr}^3\oplus \text{Od}_2\,$, given as follows:
\begin{equation}
\label{eq:TCSgluingalgebra}
    \left( \psi_t \, , \psi_\theta \, , \psi_\xi \, , B_2 \, , J^3_2 \, , A_2 \right) \longmapsto \left( -\psi_t \, , \psi_\xi \, , \psi_\theta \, , -B_2 \, , A_2 \, , J^3_2 \right) \, ,
\end{equation}
with the same action on the superpartners. This is indeed an automorphism and it can be checked that it leaves the $\SVseven$ algebra at the bottom of the diamond invariant. Applying this automorphism to $\left( \text{Fr}^1\oplus \text{Od}_3 \right)_+$ we obtain an explicit description of the remaining algebra $\left( \text{Fr}^1\oplus \text{Od}_3 \right)_-$ and its inclusion in the diamond
\begin{equation}
    J^3_{3-} = J^3_2 \, - \normord{\psi_t\,\psi_\xi} \, , \qquad A_{3-}+iB_{3-} = \, \normord{(\psi_\xi+i\psi_t)(J^3_2-iB_2)} \, .
\end{equation}
We thus conclude that the diamond associated to a TCS manifold is well-defined. We present it in \cref{fig:DiamondTCS}.

\section{Superconformal algebras for the Schoen Calabi--Yau manifold}

\label{sec:Schoen}

In this section we provide new results regarding the diamond of worldsheet algebras associated to the Schoen manifold \cite{Schoen1988}. The Schoen manifold is a Calabi--Yau 3-fold with Hodge numbers $\left( h^{1,1}, h^{2,1} \right)=(19,19)$ and it has been extensively studied together with its quotients in the context of F-theory and heterotic compactifications, see for example \cite{Morrison:1996pp, Gopakumar:1996mu, Curio:1997rn, Braun:2005ux, Braun:2005nv, Candelas:2007ac, Apruzzi:2014dza}. The Schoen manifold is also known in the literature as the split bicubic because it can be realized via a complete intersection with configuration matrix
\begin{equation}
    \left[
    \begin{array}{c|cc}
        \mathbb{CP}^1 & 1 & 1 \\
        \mathbb{CP}^2 & 3 & 0 \\
        \mathbb{CP}^2 & 0 & 3       
    \end{array}
    \right] \, .
\end{equation}
We will not need a detailed understanding of the Schoen manifold as a CICY, but it will be useful to keep in mind that the Schoen manifold can be described as a fibration over $\mathbb{CP}^1$ whose generic fiber is given by two elliptic surfaces (i.e. two 2-tori).

We are interested in the Schoen manifold as a connected sum. This was explicitly described in \cite{Braun:2017uku} using the duality between M-theory on a K3 surface and heterotic string theory on a 3-torus and we present a summary of the main ideas in \cref{sec:SchoenfromMtheory}. We also present a complementary description of the Schoen manifold as an orbifold resolution in \cref{sec:Schoenasorbifold}. We then move to a description of the worldsheet chiral algebras, introducing the corresponding diamond in \cref{sec:Schoendiamond} and discussing its genericity in \cref{sec:Schoengeneric}. We conclude in \cref{sec:Schoenautomorphisms} with a study of automorphisms of the diamond which include a realization of SYZ mirror symmetry in the worldsheet.

\subsection{The Schoen manifold as a connected sum}

\label{sec:SchoenfromMtheory}

We summarize the discussion of \cite{Braun:2017uku} leading to the connected sum decomposition of the Schoen manifold and refer the reader to the original paper for further details. The key idea is a fiberwise application of the duality between M-theory on a K3 surface and heterotic string theory on a 3-torus \cite{Acharya:1996ci, Acharya:2001gy, Gukov:2002jv}.

Consider an M-theory compactification of a TCS G$_2$-manifold\footnote{We follow the notation introduced in \cref{sec:TCSgeometry}.} and focus on its building blocks, $M_\pm=\Sc^1_{\xi\pm}\times\text{ACyl CY}_{3\pm}\,$. The standard procedure to construct ACyl Calabi--Yau 3-folds \cite{Corti:2012, Corti:2012kd} is to start from a closed K\"ahler 3-fold given by a K3 fibration over $\mathbb{CP}^1$ (satisfying certain properties) and then remove a point from the $\mathbb{CP}^1$ as well as the K3 fiber over that point. We can thus think of each $\text{ACyl CY}_{3\pm}$ as a fibration: the fiber is a K3 surface $\text{CY}_{2\pm}$ and the base $B_\pm$ is given by $\mathbb{CP}^1$ minus a point. These fibrations realize the asymptotic cylindrical behaviour \eqref{eq:TCSasymptoticbuildingblock}, where $(t_\pm,\theta_\pm)$ are the asymptotic coordinates of $B_\pm\longrightarrow\R^+_\pm\times\Sc^1_{\theta\pm}\,$.

For technical reasons, we assume that the $\text{CY}_{2\pm}$ fibers are themselves given by an elliptic fibration over $\mathbb{CP}^1$ with a well-defined section, and we further assume that the elliptic curves are holomorphically fibered over the $\mathbb{CP}^1$. These assumptions imply that when the $\text{CY}_{2\pm}$ are fibered over $B_\pm$ to give the $\text{ACyl CY}_{3\pm}\,$, the Hermitian forms $\omega_{2\pm}=\omega^I_{\pm}$ remain constant whereas the holomorphic volume forms $\Omega_{2\pm}=\omega^J_{\pm}+i\,\omega^K_{\pm}$ vary over the fibration.

We can now consider the dual heterotic picture by applying M-theory/heterotic duality to the K3 fibers $\text{CY}_{2\pm}\,$. These are then replaced by 3-torus fibers $\mathbb{T}^3_\pm$ while keeping the base $B_\pm$ and the external circle $\Sc^1_{\xi\pm}$ intact. This duality establishes a relation between the circles $\Sc^1_{a\pm}$ of the $\mathbb{T}^3_\pm$ fibers and the Hermitian forms $\omega^a_\pm$ of the $\text{CY}_{2\pm}$ via a period decomposition, where $a\in\lbrace I,J,K\rbrace$. In particular, the $\mathbb{T}^3_\pm$ fiber factorizes as $\mathbb{T}_{f\pm}\times \Sc^1_{I\pm}$ in the following way: the $\Sc^1_{I\pm}$ circle---which is related to the Hermitian forms $\omega^I_{\pm}$ that remain constant in the $\text{CY}_{2\pm}$ fibers---is trivially fibered over $B_\pm\,$, whereas the torus $\mathbb{T}_{f\pm}=\Sc^1_{J\pm}\times\Sc^1_{K\pm}$ is non-trivially fibered over $B_\pm$ and gives rise to a new open Calabi--Yau 2-fold.

Therefore, after M-theory/heterotic duality we obtain the dual building blocks $\widetilde{M}_\pm = \Sc^1_{\xi\pm}\times \Sc^1_{I\pm} \times \widetilde{\text{CY}}_{2\pm} \,$, where the $\widetilde{\text{CY}}_{2\pm}$ factors are given by the $\mathbb{T}_{f\pm}$ fibration over $B_\pm\,$. These building blocks are again asymptotically cylindrical with a cross-section given by a product of five circles:
\begin{equation}
    \widetilde{M}_\pm = \Sc^1_{\xi\pm}\times \Sc^1_{I\pm} \times \widetilde{\text{CY}}_{2\pm} \longrightarrow \R^+_\pm\times\Sc^1_{\theta\pm}\times\Sc^1_{\xi\pm}\times\Sc^1_{I\pm}\times\Sc^1_{J\pm}\times\Sc^1_{K\pm}= \R^+_\pm\times \widetilde{N}_\pm \, .
\end{equation}
In what follows we denote by $(x_{I\pm},x_{J\pm},x_{K\pm})$ the coordinates on the circles $\Sc^1_{I\pm}\times\Sc^1_{J\pm}\times\Sc^1_{K\pm}\,$. Since the glued TCS manifold has G$_2$-holonomy, the duality implies that the compact manifold $\widetilde{M}$ obtained after gluing $\widetilde{M}_\pm$ will be a Calabi--Yau 3-fold. We therefore want to define an $\text{SU}(3)$-structure on each building block $\widetilde{M}_\pm\,$, this can be done mimicking the ACyl CY asymptotic relations \eqref{eq:hermitianasymptotic} and \eqref{eq:holomorphicasymptotic}
\begin{equation}
\label{eq:SchoenSU3}
    \widetilde{\omega}_{3\pm}=\dd\xi_\pm\wedge\dd x_{I\pm} + \widetilde{\omega}_{2\pm} \, , \qquad  \widetilde{\Omega}_{3\pm}=(\dd x_{I\pm}-i\dd\xi_\pm)\wedge\widetilde{\Omega}_{2\pm} \, ,
\end{equation}
where $(\widetilde{\omega}_{2\pm} , \widetilde{\Omega}_{2\pm} )$ are the Hermitian and holomorphic volume forms on the $\widetilde{\text{CY}}_{2\pm}\,$. Their correct asymptotic behaviour is again provided by the duality, for example $\widetilde{\omega}_{2\pm,\infty}$ can be read off directly from \eqref{eq:coassociativeasymptotic} following the identification between Hermitian forms and circles,
\begin{align}
    \label{eq:CY2hermitianasymptotic}
    \widetilde{\omega}_{2\pm} & \longrightarrow \widetilde{\omega}_{2\pm,\infty}=\dd\theta_\pm\wedge\dd x_{J\pm}+\dd t_\pm\wedge\dd x_{K\pm} \, ,\\
    \label{eq:CY2holomorphicasymptotic}
    \widetilde{\Omega}_{2\pm} & \longrightarrow \widetilde{\Omega}_{2\pm,\infty}=(\dd  x_{J\pm}-i\dd\theta_\pm)\wedge(\dd x_{K\pm}-i\dd t_\pm) \, .
\end{align}
Combining \eqref{eq:CY2hermitianasymptotic} and \eqref{eq:CY2holomorphicasymptotic} with \eqref{eq:SchoenSU3} we find the asymptotic behaviour of the $\text{SU}(3)$-structure with respect to the neck region
\begin{align}
    \label{eq:CY3hermitianasymptotic}
    \widetilde{\omega}_{3\pm} & \longrightarrow \widetilde{\omega}_{3\pm,\infty}=\dd\xi_\pm\wedge\dd x_{I\pm} +\dd\theta_\pm\wedge\dd x_{J\pm}+\dd t_\pm\wedge\dd x_{K\pm} \, ,\\
    \label{eq:CY3holomorphicasymptotic}
    \widetilde{\Omega}_{3\pm} & \longrightarrow \widetilde{\Omega}_{3\pm,\infty}=(\dd x_{I\pm}-i\dd\xi_\pm)\wedge(\dd  x_{J\pm}-i\dd\theta_\pm)\wedge(\dd x_{K\pm}-i\dd t_\pm) \, .
\end{align}
We have fully described the building blocks and the neck region of the dual manifold for the heterotic compactification. It remains to identify the gluing map $\widetilde{F}$ that can be used to fuse $\widetilde{M}_\pm$ together to obtain the compact dual manifold $\widetilde{M}$.

The gluing map can be identified from the TCS gluing map $F$. The gluing of the base and the external circle are exactly as in \eqref{eq:TCSgluingmap1}, where the internal and external circles are interchanged, $\Sc^1_{\theta\pm}\leftrightarrow\Sc^1_{\xi\pm}\,$. The gluing of the $\mathbb{T}^3_\pm$ fibers is obtained from \eqref{eq:TCSgluingmap2} identifying the Hermitian forms with the corresponding circles. This shows the trivially fibered circle is exchanged with one of the $\mathbb{T}_{f\pm}$ circles, $\Sc^1_{I\pm}\leftrightarrow\Sc^1_{J\pm}\,$, whereas the leftover circles reverse orientation upon gluing. Explicitly, the action of the gluing map on the forms is given by
\begin{align}
\label{eq:Schoengluingmap1}
    \widetilde{F}^*(\dd t_-)&=-\dd t_+\, ,& \widetilde{F}^*(\dd\xi_-)&=\dd\theta_+\, ,& \widetilde{F}^*(\dd\theta_-)&=\dd\xi_+\, , \\
    \label{eq:Schoengluingmap2}
    \widetilde{F}^*(\dd x_{K-})&=-\dd x_{K+}\, ,& \widetilde{F}^*(\dd x_{I-})&=\dd x_{J+}\, ,& \widetilde{F}^*(\dd x_{J-})&=\dd x_{I+}\, .
\end{align}
It is immediate to check from \eqref{eq:CY3hermitianasymptotic} and \eqref{eq:CY3holomorphicasymptotic} that the gluing map $\widetilde{F}$ identifies the asymptotic ends of the Hermitian forms $\widetilde{\omega}_{3\pm}$ and the holomorphic volume forms $\widetilde{\Omega}_{3\pm}\,$. Thus, we obtain a well-defined $\text{SU}(3)$-structure on $\widetilde{M}$ and denote its associated forms by $(\widetilde{\omega}_3,\widetilde{\Omega}_3)$.

In the limit of large neck and after a small deformation the original TCS manifold has G$_2$-holonomy, by duality we expect that the same happens for the connected sum we have just described. Therefore, in the appropriate limit the manifold $\widetilde{M}$ obtained upon gluing is a Calabi--Yau 3-fold.

In \cite{Braun:2017uku} the authors make use of a Mayer--Vietoris sequence to compute the Hodge numbers of $\widetilde{M}$. They find $\left( h^{1,1}, h^{2,1} \right)=(19,19)$, which uniquely identifies the connected sum Calabi--Yau 3-fold to be the Schoen manifold.

An important observation is that all M-theory compactifications on TCS G$_2$-manifolds with the technical assumptions we have considered---such as having elliptically fibered K3 fibers---are dual to heterotic compactifications on the same Calabi--Yau manifold. This might seem surprising at first, however the duality also specifies a choice of vector bundles over the Schoen manifold that depends on the TCS manifold and distinguishes the dual heterotic theories \cite{Braun:2017uku}. For the purpose of this note, we just focus on the connected sum decomposition of the Schoen manifold.

\subsection{The Schoen manifold as an orbifold resolution}

\label{sec:Schoenasorbifold}

The Schoen manifold can also be described as a crepant resolution of a toroidal orbifold of the form $\mathbb{T}^6/(\mathbb{Z}_2\times\mathbb{Z}_2)$ \cite{Joyce:1996b}. In order to describe the associated diamond of algebras it will be useful to establish a connection between the toroidal description and the connected sum construction, and we do so following \cite{Braun:2017uku} again.\footnote{Note however that our conventions differ slightly from those of \cite{Braun:2017uku}.}

We parametrize $\mathbb{T}^6$ with coordinates $x^i$ ranging from $0$ to $1$, with $i=1,\dots,6$. We choose complex coordinates
\begin{equation}
\label{eq:orbifoldcoords}
    z^1=x^1+i x^2 \, , \qquad z^2=x^3+i x^4 \, , \qquad z^3=x^5+i x^6 \, ,
\end{equation}
and define the following $\text{SU}(3)$-structure on the torus\footnote{Here we are rotating the standard holomorphic volume form by a factor of $i$, this is done simply to match the conventions we have inherited from the definition of ACyl CY manifolds.}
\begin{equation}
\label{eq:orbifoldSU3structure}
    \widetilde{\omega_3}=\frac{i}{2} \left( \dd z^1 \wedge \dd \bar{z}^1 + \dd z^2 \wedge \dd \bar{z}^2 + \dd z^3 \wedge \dd \bar{z}^3 \right) \, , \qquad \widetilde{\Omega_3}=i \, \dd z^1\wedge \dd z^2 \wedge \dd z^3 \, .
\end{equation}
The quotient group $\mathbb{Z}_2\times\mathbb{Z}_2$ has two generators {$\alpha, \beta$} whose action on the coordinates can be found in \cref{tab:discreteaction} for the choice $(b_1,b_3,b_5)=(0,0,\frac{1}{2})$. The singularities of the orbifold $\mathbb{T}^6/(\mathbb{Z}_2\times\mathbb{Z}_2)$ can be resolved to obtain a Calabi--Yau manifold with Hodge numbers $\left( h^{1,1}, h^{2,1} \right)=(19,19)$, we thus find the Schoen manifold.

We can identify the connected sum decomposition even before the resolution: the orbifold coordinate $x_5$ ranges within the interval $[0,1/4]$ and we can pull the orbifold apart along this direction. The generic fiber is a product of five circles $(\Sc^1)^5$, as corresponds to the neck region, and the building blocks correspond to the ends $\widetilde{M}_+=\left((\Sc^1)^5\times \R\right) / \mathbb{Z}_2^\alpha$ at $x_5=0$ and $\widetilde{M}_-=\left((\Sc^1)^5\times \R\right) / \mathbb{Z}_2^\beta$ at $x_5=1/4$. The quotient groups at these ends act non-trivially on the $\widetilde{\text{CY}}_{2\pm}$ factors while leaving the trivial circles intact. As a result, we can identify the orbifold coordinates with their corresponding asymptotic coordinates as follows
\begin{align}
\label{eq:M+coords}
    &\widetilde{M}_+:&  x^1&=\xi_+\, , & x^2&=x_{I+} \, , & x^3&=\theta_+\, , & x^4&=x_{J+} \, , & x^5&=t_+\, , & x^6&=x_{K+} \, ,  \\
    \label{eq:M-coords}
    &\widetilde{M}_-:&  x^1&=\theta_-\, , & x^2&=x_{J-} \, , & x^3&=\xi_-\, , & x^4&=x_{I-} \, , & x^5&=-t_-\, , & x^6&=-x_{K-} \, .
\end{align}
This provides a coordinate description of the Schoen orbifold. Note in particular that the Hermitian and holomorphic volume forms from \eqref{eq:orbifoldSU3structure} agree with those of \eqref{eq:SchoenSU3} in the limits \eqref{eq:CY3hermitianasymptotic}, \eqref{eq:CY3holomorphicasymptotic} upon the identifications \eqref{eq:M+coords} and \eqref{eq:M-coords}.

{\renewcommand{\arraystretch}{1.2}
\begin{table}
\begin{tabular}{ |c||c|c|c|c|c|c| } 
\hline
   & $x^1$ & $x^2$ & $x^3$ & $x^4$ & $x^5$ & $x^6$ \\
 \hline
 \hline
 $\alpha$ & $+$ & $+$ & $-$ & $-$ & $-$ & $-$ \\
 \hline
$\beta$ & $b_1-$ & $-$ & $b_3+$ & $+$ & $b_5-$ & $-$ \\ 
\hline
\end{tabular}
\caption{Action of $\mathbb{Z}_2\times\mathbb{Z}_2$ on $\mathbb{T}^6$. The choice of parameters $(b_1,b_3,b_5)=(0,0,\frac{1}{2})$ produces the Schoen manifold after resolution. The $\pm$ entries correspond to a global $\pm$ sign whereas the $\frac{1}{2}-$ entries correspond to $x^i\mapsto-x^i+\frac{1}{2}\,$.}
\label{tab:discreteaction}
\end{table}}

We conclude with a brief comment about alternative choices of parameters $(b_1,b_3,b_5)$. Donagi and Wendland \cite{Donagi:2008xy} obtained a complete classification of Calabi--Yau 3-folds arising from orbifolds of $\mathbb{T}^6$ whose quotient group is given by an abelian extension of $\mathbb{Z}_2\times\mathbb{Z}_2\,$. For the particular case where the group is precisely $\mathbb{Z}_2\times\mathbb{Z}_2\,$, in addition to the Schoen manifold we also have the choices $(0,0,0)$, $(0,\frac{1}{2},\frac{1}{2})$ and $(\frac{1}{2},\frac{1}{2},\frac{1}{2})$.\footnote{The first one is the Vafa--Witten orbifold \cite{Vafa:1994rv} and the second is a Borcea--Voisin manifold \cite{Borcea:1996mxz, Voisin:246106} with Hodge numbers $\left( h^{1,1}, h^{2,1} \right)=(11,11)$ \cite{Ferrara:1995yx}.} We point out how the last two cases can be pulled apart along the coordinates $x^5$, $x^3$ and $x^5$, $x^3$, $x^1$. This suggests that these manifolds might also admit a connected sum description.

\subsection{Chiral algebra perspective}

\label{sec:Schoendiamond}

In this section we describe the diamond of chiral algebras corresponding to the connected sum description of the Schoen manifold. The procedure is similar to that of \cref{sec:TCSalgebra}: the geometry of the manifolds determines the corresponding chiral algebras, and the asymptotic relations between the forms on these manifolds provide a natural ansatz for the algebra inclusions in the diamond. We have verified that the ans\"atze we present indeed define the desired algebra inclusions.\footnote{All the algebra computations in this note have been performed with the help of the package \emph{OPEdefs} by Thielemans \cite{Thielemans:1994er}.} This involves a careful treatment of the Odake null fields, see \eqref{eq:nullfields}.

As motivated in sections~\ref{sec:introAlgebras} and \ref{sec:introConnSumAlgebras}, we will associate an Odake algebra $\text{Od}_n$ to each Calabi--Yau $n$-fold and a Free algebra $\text{Fr}^1$ to each $\Sc^1$ or $\R$. We will describe inclusions of the Odake algebra through the explicit expressions of the operators $(G_n,J^3_n,A_n,B_n)$, this is enough since the expressions of the remaining operators are implied by the OPEs.\footnote{The explicit OPEs of all the operators involved can be found in the appendices of \cite{Galdeano:2022mki}.}

The neck region of the Schoen manifold is of the form $\R^+\times\left(\Sc^1\right)^5$, so the associated algebra is simply $\text{Fr}^6$. We denote the Free algebras $\text{Fr}^1_i$ and their operators by $(\psi_i,j_i)$ with $i=1,\dots,6$, matching the coordinates of the orbifold description \eqref{eq:orbifoldcoords}.

The building blocks are of the form $\widetilde{M}_\pm = \Sc^1_{\xi\pm}\times \Sc^1_{I\pm} \times \widetilde{\text{CY}}_{2\pm}$  so the corresponding algebras on the lateral tips of the diamond are $\left( \text{Fr}^2\oplus \text{Od}_2 \right)_\pm \,$. Each Free algebra is identified with a Free algebra in $\text{Fr}^6$ via the coordinate identifications \eqref{eq:M+coords} and \eqref{eq:M-coords}. The $\text{Od}_2$ factors are given by a free field realization motivated by the asymptotic behaviour of the Hermitian and holomorphic volume forms \eqref{eq:CY2hermitianasymptotic}, \eqref{eq:CY2holomorphicasymptotic}, together with \eqref{eq:M+coords} and \eqref{eq:M-coords} again. For $\left( \text{Fr}^2\oplus \text{Od}_2 \right)_+\subset\text{Fr}^6$ we have
\begin{align}
\begin{split}
\label{eq:Od2+inFree}
    \text{Fr}^1_{\xi+}&=\text{Fr}^1_1 \, , \qquad \text{Fr}^1_{I+}=\text{Fr}^1_2 \, , \\
     G_{2+}&=G^3 + G^4 + G^5 + G^6 \, ,\\
     J^3_{2+}&=\,\normord{\psi_3\psi_4}+\normord{\psi_5\psi_6} \, ,\\
    A_{2+}+i B_{2+}&=\,\normord{(\psi_4-i\psi_3)(\psi_6-i\psi_5)} \, ,
\end{split}
\end{align}
whereas for $\left( \text{Fr}^2\oplus \text{Od}_2 \right)_-\subset\text{Fr}^6$ we obtain
\begin{align}
\begin{split}
\label{eq:Od2-inFree}
    \text{Fr}^1_{\xi-}&=\text{Fr}^1_3 \, , \qquad \text{Fr}^1_{I-}=\text{Fr}^1_4 \, , \\
     G_{2-}&=G^1 + G^2 + G^5 + G^6 \, ,\\
     J^3_{2-}&=\,\normord{\psi_1\psi_2}+\normord{\psi_5\psi_6} \, ,\\
    A_{2-}+i B_{2-}&=\,\normord{(\psi_2-i\psi_1)(-\psi_6+i\psi_5)} \, .
\end{split}
\end{align}
The algebra at the bottom of the diamond must be $\text{Od}_3$ since the Schoen manifold is a Calabi--Yau 3-fold. It is well-known that $\text{Od}_3$ can be embedded into $\text{Fr}^2\oplus \text{Od}_2\,$, recall \eqref{eq:Od3fromOd2} and see \cite{Fiset:2018huv}. The precise embedding in our case is motivated by the $\text{SU}(3)$-structure on the building blocks \eqref{eq:SchoenSU3}. For $\text{Od}_3\subset\left( \text{Fr}^2\oplus \text{Od}_2 \right)_+$ we find
\begin{align}
\begin{split}
\label{eq:Od3inOd2+}
     G_{3}&=G^1 + G^2 + G_{2+} \, ,\\
     J^3_{3}&=\,\normord{\psi_1\psi_2}+J^3_{2+} \, ,\\
    A_{3}+i B_{3}&=\,\normord{(\psi_2-i\psi_1)\left(A_{2+}+i B_{2+}\right)} \, ,
\end{split}
\end{align}
whereas for the embedding $\text{Od}_3\subset\left( \text{Fr}^2\oplus \text{Od}_2 \right)_-$ we have
\begin{align}
\begin{split}
\label{eq:Od3inOd2-}
     G_{3}&=G^3 + G^4 + G_{2-} \, ,\\
     J^3_{3}&=\,\normord{\psi_3\psi_4}+J^3_{2-} \, ,\\
    A_{3}+i B_{3}&=\,\normord{(\psi_4-i\psi_3)\left(A_{2-}+i B_{2-}\right)} \, .
    \end{split}
\end{align}
Combining \eqref{eq:Od3inOd2+} with \eqref{eq:Od2+inFree} and \eqref{eq:Od3inOd2-} with \eqref{eq:Od2-inFree} we obtain the same free field realization of $\text{Od}_3$ inside $\text{Fr}^6$, showing that the diamond closes in the way of \cref{fig:DiamondSchoen}. Furthermore, this realization agrees with the ansatz suggested by the $\text{SU}(3)$-structure on the toroidal orbifold \eqref{eq:orbifoldSU3structure}
\begin{align}
\begin{split}
\label{eq:Od3inFree}
     G_{3}&=G^1 + G^2 + G^3 + G^4 + G^5 + G^6 \, ,\\
     J^3_{3}&=\,\normord{\psi_1\psi_2}+\normord{\psi_3\psi_4}+\normord{\psi_5\psi_6} \, ,\\
    A_{3}+i B_{3}&=\,\normord{(\psi_2-i\psi_1)(\psi_4-i\psi_3)(\psi_6-i\psi_5)} \, .
\end{split}
\end{align}
Finally, we can study the automorphism of the diamond corresponding to the gluing map \eqref{eq:Schoengluingmap1}, \eqref{eq:Schoengluingmap2}. Identifying the coordinates of the orbifold with those of $\widetilde{M}_+$ via \eqref{eq:M+coords}, we obtain
\begin{align}
\label{eq:Schoengluingauto1}
    \left(\psi_5,j_5\right)&\longmapsto \left(-\psi_5,-j_5\right)\, ,& \left(\psi_3,j_3\right)&\longmapsto \left(\psi_1,j_1\right)\, ,& \left(\psi_1,j_1\right)&\longmapsto \left(\psi_3,j_3\right)\, , \\
    \label{eq:Schoengluingauto2}
    \left(\psi_6,j_6\right)&\longmapsto \left(-\psi_6,-j_6\right)\, ,& \left(\psi_4,j_4\right)&\longmapsto \left(\psi_2,j_2\right)\, ,& \left(\psi_2,j_2\right)&\longmapsto \left(\psi_4,j_4\right)\,  .
\end{align}
This is an automorphism of $\text{Fr}^6$ that acts as the identity on the bottom algebra $\text{Od}_3$ and maps $\left( \text{Fr}^2\oplus \text{Od}_2 \right)_+$ to $\left( \text{Fr}^2\oplus \text{Od}_2 \right)_-$ and vice-versa, as can be seen from \eqref{eq:Od3inFree}, \eqref{eq:Od2+inFree} and \eqref{eq:Od2-inFree}.

The structure of the diamond and the gluing automorphism are analogous to those of the TCS discussed in \cref{sec:TCSalgebra}. This should not come as a surprise since the connected sum structure of the Schoen manifold is obtained from that of the TCS manifold via M-theory/heterotic duality.

\begin{figure}
\begin{tikzpicture}
\node at (2.8,0.55) {\rotatebox{240}{$\subset$}};
\node at (4.2,0.55) {\rotatebox{300}{$\subset$}};
\node at (2.8,-0.55) {\rotatebox{300}{$\subset$}};
\node at (4.2,-0.55) {\rotatebox{240}{$\subset$}};
\node at (3.5,1) {$\R^+\times\left(\Sc^1\right)^5$};
\node at (1.7,0) {$\Sc^1_{\xi+}\times \Sc^1_{I+} \times \widetilde{\text{CY}}_{2+}$};
\node at (5.3,0) {$\Sc^1_{\xi-}\times \Sc^1_{I-} \times \widetilde{\text{CY}}_{2-}$};
\node at (3.5,-1) {$\widetilde{\text{CY}}_3$};
\node at (7.2,0) {$\Leftrightarrow$};
\node at (9.3,0.55) {\rotatebox{60}{$\subset$}};
\node at (10.7,0.55) {\rotatebox{120}{$\subset$}};
\node at (9.3,-0.55) {\rotatebox{120}{$\subset$}};
\node at (10.7,-0.55) {\rotatebox{60}{$\subset$}};
\node at (10,1) {$\text{Fr}^6$};
\node at (8.8,0) {$\left( \text{Fr}^2\oplus \text{Od}_2 \right)_+$};
\node at (11.3,0) {$\left( \text{Fr}^2\oplus \text{Od}_2 \right)_-$};
\node at (10,-1) {$\text{Od}_3$};
\end{tikzpicture}
\caption{Diamond of submanifold embeddings and diamond of subalgebra inclusions for the Schoen manifold.}
\label{fig:DiamondSchoen}
\end{figure}

\subsection{Is the construction generic?}

\label{sec:Schoengeneric}

We now address the relationship between the algebra $\text{Od}_3$ at the bottom of the diamond and the intersection $\left( \text{Fr}^2\oplus \text{Od}_2 \right)_+ \cap\left( \text{Fr}^2\oplus \text{Od}_2 \right)_-$ of the algebras at the lateral tips. We expect these two algebras to be equal. The reason is that this worldsheet algebra statement would correspond in geometry to the statement that the gluing procedure of \cref{sec:SchoenfromMtheory} produces a manifold with holonomy precisely equal to $\text{SU}(3)$---and we know this is indeed the case for the Schoen manifold.

We know from the diamond that $\text{Od}_3\subset \left( \text{Fr}^2\oplus \text{Od}_2 \right)_+ \cap\left( \text{Fr}^2\oplus \text{Od}_2 \right)_-\,$, so the equality could be verified by checking that both algebras have the same number of independent operators of each conformal weight. Equivalently, we could show the agreement between the vacuum module characters of both chiral algebras, defined as
\begin{equation}
    \chi(\mathcal{A})=\tr(q^{L_0-\frac{c}{24}}) \, ,
\end{equation}
where the trace is taken over the vacuum module and $q=e^{2\pi i\tau}$. The character of the Odake algebra with $n=3$ can be found in \cite{Odake:1989dm},
\begin{equation}
\label{eq:Od3character}
    \chi(\text{Od}_3)=\sum_{m\in\mathbb{Z}} \frac{\left(1-q\right)  q^{m^2+m-\frac{1}{2}-\frac{9}{24}}}{\left( 1+q^{m-\frac{1}{2}} \right)\left( 1+q^{m+\frac{1}{2}} \right)} \prod_{n\in\mathbb{N}}\frac{\left( 1+q^{n-\frac{1}{2}} \right)^2}{\left( 1-q^n \right)^2} \, .
\end{equation}
The character of the algebras $\left( \text{Fr}^2\oplus \text{Od}_2 \right)_\pm$ is also well-known, since $\text{Od}_2$ is isomorphic to the $\mathcal{N}=4$ Virasoro algebra whose character can be found in \cite{Eguchi:1987wf}. Unfortunately, we do not have an analytic expression for the character of the intersection algebra $\left( \text{Fr}^2\oplus \text{Od}_2 \right)_+ \cap\left( \text{Fr}^2\oplus \text{Od}_2 \right)_-\, $. Therefore, we can only verify our conjecture numerically.

Expanding the character $\chi$ in powers of $q$, we can read off the number of independent fields at level $h$ as the coefficient of the $q^{h-\frac{c}{24}}$ term. From \eqref{eq:Od3character} we find
\begin{equation}
    \chi(\text{Od}_3)=q^{-\frac{3}{8}}\left( 1 + q + 4 q^{3/2} + 5 q^2 + 6 q^{5/2} +  10 q^3 + 18 q^{7/2} + 27 q^4 + 36 q^{9/2} + \dots  \right).
\end{equation}
We have verified that the number of independent fields of the intersection algebra agrees with those of $\text{Od}_3$ up to level $9/2$, providing support to our conjecture.

\subsection{Automorphisms and mirror symmetry}

\label{sec:Schoenautomorphisms}

In \cref{sec:SchoenfromMtheory} we have described the Schoen manifold as a $\mathbb{T}^3$ fibration. It is well-known this actually constitutes an SYZ fibration \cite{Strominger:1996it} and we can obtain a mirror symmetry map by performing three T-dualities along the circles in the fiber. As the Schoen manifold is self-dual \cite{Hosono:1997hp}, the mirror map in this case maps the Schoen manifold to itself.

Mirror symmetry can also be described from the worldsheet algebra perspective. Two mirror manifolds provide equivalent superconformal field theories upon compactification, so the underlying chiral algebras must remain the same. Therefore, a mirror map in geometry corresponds to an automorphism of the associated W-algebra.

For the Odake algebra, the \emph{Mirror symmetry} automorphism \textbf{M} is given by
\begin{equation}
    \textbf{M}:\,\left(J^3_n,G^3_n,B_n,D_n\right)\longmapsto \left(-J^3_n,-G^3_n,-B_n,-D_n\right)\, .
\end{equation}
We will also need the \emph{Phase} automorphism $\textbf{Ph}^\pi$, that has the geometrical interpretation of a rotation of the holomorphic volume form by an angle $\pi$,\footnote{We will encounter both $\textbf{M}$ and $\textbf{M}\circ\textbf{Ph}^\pi$ as the automorphisms associated to mirror maps. The appearance of the $\textbf{Ph}^\pi$ factor depends on the dimension of the Calabi--Yau manifold and whether the supersymmetric cycle is calibrated by $\Re(\Omega_n)$ or $\Im(\Omega_n)$.}
\begin{equation}
    \textbf{Ph}^\pi:\,\left(A_n,B_n,C_n,D_n\right)\longmapsto \left(-A_n,-B_n,-C_n,-D_n\right)\, .
\end{equation}
Similarly, performing a T-duality along a circle has a corresponding automorphism in the associated Free algebra, that we also call the \emph{T-duality} automorphism \textbf{T},
\begin{equation}
    \textbf{T}:\,\left(\psi,j\right)\longmapsto \left(-\psi,-j\right)\, .
\end{equation}
We now want to consider the mirror map of the Schoen manifold $\widetilde{M}$ obtained by T-dualising the $\mathbb{T}^3$ fiber. Since $\widetilde{M}$ is constructed by gluing two building blocks $\widetilde{M}_\pm$ with a $\mathbb{T}^3$ fibration, the $\widetilde{M}$ mirror map should give rise to a mirror map for each of the building blocks $\widetilde{M}_\pm\,$. Furthermore, these maps must agree over the neck region since the T-duality can be performed consistently over the whole $\widetilde{M}$.

At the level of the chiral algebras, this implies that the mirror map must be an automorphism of the whole diamond of algebras associated to the Schoen manifold. Therefore, we must be able to describe the mirror map by an automorphism of the top algebra of the diamond that reduces to automorphisms of the algebras at the lateral tips and the bottom of the diamond.

This automorphism is easily found from the orbifold perspective, where the top algebra $\text{Fr}^6$ can be directly associated to the orbifold coordinates. In our notation, the $\mathbb{T}^3$ fiber corresponds to the orbifold coordinates $(x^2,x^4,x^6)$. Therefore, we define the mirror automorphism by a composition of three \textbf{T} maps along the corresponding Free algebras: $\textbf{T}_2\circ\textbf{T}_4\circ\textbf{T}_6\,$.

We now study the restriction of this automorphism to the subalgebras of the diamond. For the algebras in the lateral tips $\left( \text{Fr}^2\oplus \text{Od}_2 \right)_\pm\,$, we obtain a \textbf{T} map along the Free algebras associated to the circles $\Sc^1_{I\pm}$ and a \textbf{M} map on the $\text{Od}_2\,$. For the bottom $\text{Od}_3$ algebra, we obtain the automorphism $\textbf{M}\circ \textbf{Ph}^\pi$.

Therefore, $\textbf{T}_2\circ\textbf{T}_4\circ\textbf{T}_6$ provides an automorphism of the diamond that reduces to mirror symmetry automorphisms for the different Odake algebras, reproducing the SYZ mirror map at the level of chiral algebras.

These automorphisms can be understood geometrically via the orbifold picture, as they arise from T-dualities along toroidal supersymmetric fibers. Indeed, it is not hard to see from \eqref{eq:CY2holomorphicasymptotic} together with \eqref{eq:M+coords} and \eqref{eq:M-coords} that the tori $\Sc^1_4\times\Sc^1_6$ and $\Sc^1_2\times\Sc^1_6$ are calibrated by $\Re( \widetilde{\Omega}_{2+})$ and $\Re( \widetilde{\Omega}_{2-})$, respectively. Therefore, these tori constitute special Lagrangian submanifolds of $\widetilde{\text{CY}}_{2\pm}$ \cite{Joyce:2007} and T-duality should give rise to a mirror automorphism in $\text{Od}_2\,$. Analogously, the $\mathbb{T}^3$ fiber is calibrated by $\Re(\widetilde{\Omega}_3)$, see \eqref{eq:orbifoldSU3structure}, so it is a special Lagrangian submanifold of $\widetilde{M}$ as expected.

Finally, we discussed in \cref{sec:Schoenasorbifold} the existence of other Calabi--Yau orbifolds that can be pulled along different coordinates. If these orbifolds do admit connected sum structures, their associated diamonds of chiral algebras will be the same as the one of the Schoen manifold. This motivates a more general study of the automorphisms of our diamond, as these orbifolds could admit supersymmetric fibrations along directions different from the ones we have just presented. A similar analysis was performed for $\mathbb{T}^8$ orbifolds describing GCS Spin(7)-manifolds in \cite{Fiset:2021ruv}.

We investigate which combinations of $\textbf{T}_i$ maps preserve the diamond of \cref{fig:DiamondSchoen} and descend to automorphisms of the subalgebras. We find that the allowed combinations are determined by the terms in the holomorphic volume form $\widetilde{\Omega}_3\,$, which can be read off \eqref{eq:orbifoldSU3structure}. Geometrically, these correspond to 3-tori of $\mathbb{T}^6$ calibrated by either $\Re(\widetilde{\Omega}_3)$ or $\Im(\widetilde{\Omega}_3)$. We have collected the automorphisms and their action on the diamond in tables~\ref{tab:automorphisms1} and \ref{tab:automorphisms2}.

We find that, in addition to the automorphism associated to the SYZ fibration of the Schoen manifold, there are seven additional potential mirror maps. It would be interesting to explicitly check which ones are realized by supersymmetric fibrations of the Calabi--Yau orbifolds from \cite{Donagi:2008xy}.

{\renewcommand{\arraystretch}{1.2}
\begin{table}
\begin{tabular}{ |c|c|c|c|c| } 
\hline
 $\text{Coordinates}$  & $(2,4,6)$ & $(2,3,5)$ & $(1,4,5)$ & $(1,3,6)$ \\
 \hline
 \hline
 $\text{Fr}^6$ & $\textbf{T}_2\circ\textbf{T}_4\circ\textbf{T}_6$ & $\textbf{T}_2\circ\textbf{T}_3\circ\textbf{T}_5$ & $\textbf{T}_1\circ\textbf{T}_4\circ\textbf{T}_5$ & $\textbf{T}_1\circ\textbf{T}_3\circ\textbf{T}_6$ \\
 \hline
$\left( \text{Fr}^2\oplus \text{Od}_2 \right)_+$ & $\textbf{T}_2\circ\textbf{M}$ & $\textbf{T}_2\circ\textbf{M}$ & $\textbf{T}_1\circ\textbf{M}\circ \textbf{Ph}^\pi$ & $\textbf{T}_1\circ\textbf{M}\circ \textbf{Ph}^\pi$ \\ 
\hline
$\left( \text{Fr}^2\oplus \text{Od}_2 \right)_-$ & $\textbf{T}_4\circ\textbf{M}$ & $\textbf{T}_3\circ\textbf{M}\circ \textbf{Ph}^\pi$ & $\textbf{T}_4\circ\textbf{M}$ & $\textbf{T}_3\circ\textbf{M}\circ \textbf{Ph}^\pi$ \\ 
\hline
$\text{Od}_3$ & $\textbf{M}\circ \textbf{Ph}^\pi$ & $\textbf{M}\circ \textbf{Ph}^\pi$ & $\textbf{M}\circ \textbf{Ph}^\pi$ & $\textbf{M}\circ \textbf{Ph}^\pi$  \\
\hline
\end{tabular}
\caption{Automorphisms of the diamond restricting to $\textbf{M}\circ \textbf{Ph}^\pi$ in $\text{Od}_3\,$.}
\label{tab:automorphisms1}

\begin{tabular}{ |c|c|c|c|c| } 
\hline
 $\text{Coordinates}$  & $(2,4,5)$ & $(2,3,6)$ & $(1,4,6)$ & $(1,3,5)$ \\
 \hline
 \hline
 $\text{Fr}^6$ & $\textbf{T}_2\circ\textbf{T}_4\circ\textbf{T}_5$ & $\textbf{T}_2\circ\textbf{T}_3\circ\textbf{T}_6$ & $\textbf{T}_1\circ\textbf{T}_4\circ\textbf{T}_6$ & $\textbf{T}_1\circ\textbf{T}_3\circ\textbf{T}_5$ \\
 \hline
$\left( \text{Fr}^2\oplus \text{Od}_2 \right)_+$ & $\textbf{T}_2\circ\textbf{M}\circ \textbf{Ph}^\pi$ & $\textbf{T}_2\circ\textbf{M}\circ \textbf{Ph}^\pi$ & $\textbf{T}_1\circ\textbf{M}$ & $\textbf{T}_1\circ\textbf{M}$ \\ 
\hline
$\left( \text{Fr}^2\oplus \text{Od}_2 \right)_-$ & $\textbf{T}_4\circ\textbf{M}\circ \textbf{Ph}^\pi$ & $\textbf{T}_3\circ\textbf{M}$ & $\textbf{T}_4\circ\textbf{M}\circ \textbf{Ph}^\pi$ & $\textbf{T}_3\circ\textbf{M}$ \\ 
\hline
$\text{Od}_3$ & $\textbf{M}$ & $\textbf{M}$ & $\textbf{M}$ & $\textbf{M}$  \\
\hline
\end{tabular}
\caption{Automorphisms of the diamond restricting to $\textbf{M}$ in $\text{Od}_3\,$.}
\label{tab:automorphisms2}
\end{table}}

\section{Conclusion}

\label{sec:Conclusion}

In this note we have discussed connected sum manifolds from the point of view of their associated superconformal algebras. After introducing some necessary background, we have revisited the proposal of \cite{Fiset:2021ruv} that assigns a unique diamond of algebra inclusions to a connected sum manifold. This diamond reflects the piecewise structure of the manifold and provides a counterpart to the geometry at the level of chiral algebras.

Then, we have summarized the case of TCS G$_2$-manifolds---discussed in \cite{Fiset:2018huv, Fiset:2021ruv} in a different format---before presenting new evidence for the proposal. In particular, we have described in detail the realization of the diamond associated to the Schoen Calabi--Yau manifold, studying different algebra inclusions and analysing its automorphisms. To this end, we have used the connected sum decomposition first described in \cite{Braun:2017uku}.

There are several future directions that would prove interesting to pursue. First of all, it would be illuminating to test the diamond proposal on additional manifolds admitting a connected sum decomposition. The ``doubling constructions'' of Doi and Yotsutani \cite{Doi:2013, Doi:2015, Doi:2019} would be natural candidates. Note that for the case of Calabi--Yau 3-folds \cite{Doi:2013} this construction recovers some of the Borcea--Voisin manifolds \cite{Borcea:1996mxz, Voisin:246106}---including the Schoen manifold---so a first step would be verifying if the Schoen diamond extends to this more general setting.

We could also follow the ideas of \cref{sec:SchoenfromMtheory} and apply different string dualities to known compactifications on connected sum manifolds. In this way we could obtain new descriptions of connected sum manifolds and their associated diamonds. For example, one could try to study the Calabi--Yau 4-fold dual to a TCS manifold via M-theory/F-theory duality \cite{Braun:2017uku}.

Testing the proposal on manifolds beyond connected sums is another enticing possibility. By the arguments of \cref{sec:introConnSumAlgebras}, for a manifold described by several open patches it should still be possible to identify uniquely its underlying chiral algebra in terms of the algebras associated to the open patches.

On a different note, our analysis in \cref{sec:Schoenautomorphisms} shows several candidates for mirror maps of Calabi--Yau 3-folds obtained from toroidal orbifolds \cite{Donagi:2008xy}. At the moment, it is an open question whether these are realised in geometry or not. 

Finally, it would be extremely interesting to relate the discussion in this note to the chiral de Rham complex \cite{Malikov:1998dw}. This would allow us to make a connection between the vast literature in the subject and our diamond proposal, perhaps shedding some new light on our results.

\subsection*{Acknowledgements}

I would like to thank my collaborator Marc-Antoine Fiset for introducing me into this line of research and for his contributions to some of the results presented here. I would also like to thank Xenia de la Ossa and Enrico Marchetto for helpful comments and discussions. Finally, I am very grateful to the organisers of String Math 2022 for putting together a fantastic conference and for the opportunity to write this contribution.

\bibliographystyle{amsplain}

\providecommand{\bysame}{\leavevmode\hbox to3em{\hrulefill}\thinspace}
\providecommand{\MR}{\relax\ifhmode\unskip\space\fi MR }
\providecommand{\MRhref}[2]{%
  \href{http://www.ams.org/mathscinet-getitem?mr=#1}{#2}
}
\providecommand{\href}[2]{#2}

\end{document}